\documentclass[pra,aps,twocolumn,nofootinbib]{revtex4}

\usepackage{amsfonts,amssymb,amsmath,amsthm,graphicx} 
\usepackage{url}

\newtheorem{theorem}{Theorem}
\newtheorem{proposition}{Proposition}

\newtheorem{conjecture}{Conjecture}
\newtheorem{lemma}[theorem]{Lemma}

\newcommand*{\picscale}{0.5}

\newcommand*{\bbN}{\mathbb{N}}
\newcommand*{\bbR}{\mathbb{R}}

\newcommand*{\cB}{\mathcal{B}}

\newcommand*{\cE}{\mathcal{E}}

\newcommand*{\cH}{\mathcal{H}}

\newcommand*{\cK}{\mathcal{K}}

\newcommand*{\cS}{\mathcal{S}}
\newcommand*{\cU}{\mathcal{U}}
\newcommand*{\cT}{\mathcal{T}}
\newcommand*{\cV}{\mathcal{V}}

\newcommand*{\id}{\mathrm{id}}

\newcommand*{\tr}{\mathrm{tr}}
\newcommand*{\ket}[1]{| #1 \rangle}
\newcommand*{\bra}[1]{\langle #1 |}

\newcommand*{\proj}[1]{\ket{#1}\bra{#1}}

\newcommand*{\sign}{\mathrm{sign}}

\newcommand*{\eps}{\varepsilon}
\newcommand*{\Sym}{\mathrm{Sym}}

\newcommand*{\rhob}{\bar{\rho}}

\newcommand*{\sub}{\cS}

\begin{document} 

\title{Symmetry implies independence}

\author{Renato Renner}

\affiliation{Department of Applied Mathematics and Theoretical Physics
  \\ University of Cambridge, UK \\{\tt r.renner@damtp.cam.ac.uk}}

\begin{abstract}
  Given a quantum system consisting of many parts, we show that
  \emph{symmetry} of the system's state, i.e., invariance under
  swappings of the subsystems, implies that almost all of its parts
  are virtually \emph{identical} and \emph{independent} of each other.
  This result generalises de Finetti's classical representation
  theorem for infinitely exchangeable sequences of random variables as
  well as its quantum-mechanical analogue. It has applications in
  various areas of physics as well as information theory and
  cryptography. For example, in experimental physics, one typically
  collects data by running a certain experiment many times, assuming
  that the individual runs are mutually independent.  Our result can
  be used to justify this assumption.
\end{abstract}

\maketitle

\section{Introduction}

In physics, properties of a large system (e.g., the universe) are
typically inferred based on observations restricted to a small part of
it (namely the part which is accessible to our experiments).  For
example, based on experiments in a laboratory showing that a hydrogen
atom absorbs radiation at a certain wavelength, we naturally
conjecture that the same is true for all hydrogen atoms in the
universe.  In other words, we expect that a limited number of local
experiments is sufficient to derive general physical laws.

\begin{figure} 
  \centering
  \includegraphics[scale=\picscale]{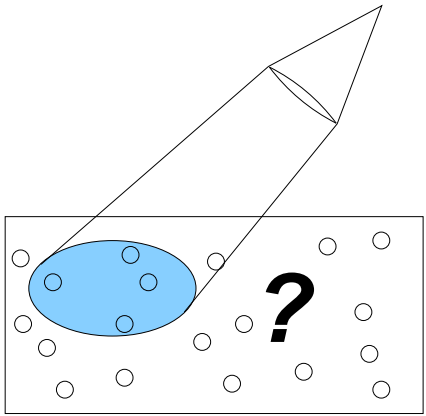}
  \caption{Given a system consisting of many subsystems (indicated by small
    circles), the goal is to infer the system's characteristics based
    on observations of only a small part of it (oval set).
  \label{fig:prob}}
\end{figure}

While this paradigm is crucial for the interpretation of experimental
data, it is, however, generally impossible to provide experimental
evidence in support of the paradigm itself. So, how else can it be
justified?  What exactly are the underlying assumptions? To answer
these questions, we consider an abstract problem, in the following
referred to as the \emph{tomography problem} (cf.\ 
Fig.~\ref{fig:prob}). Let $\sub_1, \ldots, \sub_N$ be $N$ subsystems
of a large composite system and assume that individual experiments are
performed on $k$ of the subsystems, $\sub_1, \ldots, \sub_k$, for $k
\ll N$.  The goal is to infer the physical state of the remaining
$N-k$ subsystems $\sub_{k+1}, \ldots, \sub_N$, based on this
experimental data. Note that the characteristics of the observed
subsystems $\sub_1, \ldots, \sub_k$ might, in general, be completely
unrelated to the characteristics of $\sub_{k+1}, \ldots, \sub_N$, in
which case the observation of the former does not give any information
on the latter.  Hence, in order to achieve the above goal, one needs
to make certain minimal assumptions on the structural properties of
the overall system.

In this article, we demonstrate that, for non-relativistic quantum
systems, the tomography problem can be solved under the sole
assumption that the overall system is symmetric under permutations of
the $N$ subsystems. More generally, we show that any symmetric system
can be analysed in the same way as if its subsystems were independent
and identical copies of each other|symmetry is thus sufficient to
justify the paradigm of experimental physics described at the
beginning. Remarkably, symmetry of realistic systems often holds in
general because of certain natural properties such as the
indistinguishability of identical particles.  The result thus has a
wide range of applications. These include quantum information theory
and cryptography, where it enables the generalisation of statements
which previously have only been known to be true under certain
independence assumptions.


\section{Independence and symmetry}

The physical state $\rho^N$ of an $N$-partite system is said to be
\emph{independent and identically distributed (i.i.d.)} if its $N$
parts are identical copies of some \emph{prototype state} $\sigma$,
i.e., formally, $\rho^N = \sigma^{\otimes
  N}$.\footnote{\label{ft:class}We adopt the density operator
  formalism which is commonly used in quantum mechanics. Note that the
  formalism also applies to purely classical systems. In this case,
  all density operators are diagonal with respect to the same basis
  and can be interpreted as probability distributions.} Note that, by
applying only individual measurements on a certain (sufficiently
large) number of subsystems, the corresponding prototype $\sigma$ can
be estimated to any desired accuracy.  The tomography problem
described above (cf.\ Fig.~\ref{fig:prob}) can thus be solved under
the assumption that the state of the system is i.i.d.  This
assumption, however, is mostly impossible to justify for realistic
systems.  In particular, there is no experiment on $n$ subsystems
providing enough data to exclude the possibility that there exist
correlations involving $N > n$ subsystems (see also
Example~\ref{ex:loc} below).

\begin{figure} 
  \centering
  \includegraphics[scale=\picscale]{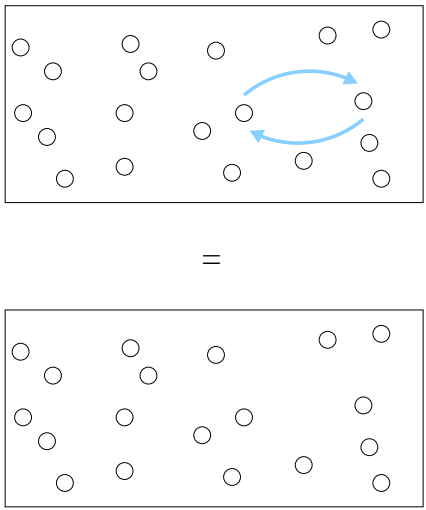}
  \caption{If the subsystems of a multi-partite system are indistinguishable then its state is symmetric, i.e., invariant under reordering of the subsystems.
  \label{fig:symm}}
\end{figure}

The state $\rho^N$ of an $N$-partite system is called \emph{symmetric}
if it is invariant under swappings of its subsystems, i.e., formally
$\pi \rho^N \pi^\dagger = \rho^N$, where $\pi$ is an arbitrary
permutation (cf.\ Fig.~\ref{fig:symm}).  This is equivalent to say
that the order in which the subsystems are represented mathematically
is independent of their physical properties. Note that any i.i.d.\ 
state is symmetric, whereas the opposite implication does generally
not hold. Moreover, for realistic systems, symmetry often follows from
certain natural properties such as the indistinguishability of its
subsystems.  Finally, in practical applications, symmetry can
sometimes be \emph{enforced} by randomly permuting the subsystems (as
illustrated below).

\section{Relation between symmetry and independence}

As discussed above, the i.i.d.\ property is strong enough to enable
applications such as tomography. However, for real physical systems,
it is often only possible to justify symmetry. This raises the
question whether symmetry of a physical state still implies a certain
similarity to i.i.d.\ states.  The Italian mathematician Bruno de
Finetti was the first to study this question for the case of classical
probabilistic systems~\cite{deFine37,deFine93}.\footnote{More
  precisely, de Finetti's theorem is formulated for probability
  distributions of random values. Note that probability distributions
  are the classical counterparts of density operators in quantum
  mechanics, i.e., they are representations of a system's state (see
  also Footnote~\ref{ft:class}).} In its generalised form \emph{de
  Finetti's representation theorem} states the
following~\cite{DiaFre80}: If the state $\rho^N$ of a classical
$N$-partite system is symmetric, then the state $\rho^n$ of any
$n$-partite subsystem, for $n \ll N$, is approximated by a
probabilistic mixture of i.i.d.\ states $\sigma^{\otimes
  n}$.\footnote{De Finetti's original work was concerned with the
  special case where $n$ is fixed and $N \to \infty$~\cite{deFine37}.}
Note that, physically, this probabilistic mixture can be interpreted
as \emph{one single} i.i.d.\ state $\sigma^{\otimes n}$ whose
prototype $\sigma$ is unknown.

\begin{figure} 
  \centering
  \includegraphics[scale=\picscale]{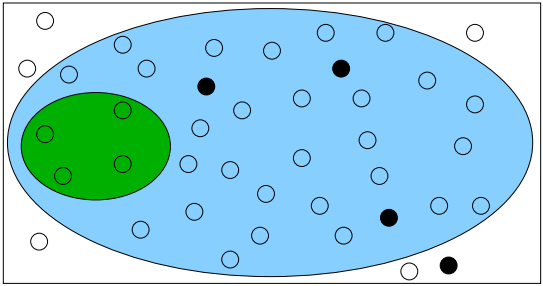}
  \caption{According to de Finetti's original representation theorem (and its
    quantum-mechanical analogs), any small part (small oval set) of a
    large symmetric system satisfies the i.i.d.\ property.  The global
    representation theorem presented here extends this statement to a
    set which almost covers the overall system (large oval set), but
    the i.i.d.\ property is slightly relaxed in that a small fraction
    of the subsystems might be in an arbitrary state (black circles).
  \label{fig:deFin}}
\end{figure}

Later, de Finetti's representation theorem was extended to quantum
theory~\cite{Stoermer69,HudMoo76,FaLeVe88,RagWer89,Petz90,CaFuSh02}.
In particular, it has been shown that the statement above holds for
any quantum system with finite-dimensional
subsystems~\cite{KoeRen05,CKMR06} as well as for certain systems with
infinite-dimensional subsystems~\cite{DeOsSc06}.  Furthermore, some of
these results have been transformed via Choi-Jamio\l{}\-kowski
isomorphism into statements about completely positive maps (CPMs),
which are used to characterise a system's dynamics~\cite{FuScSc04}.
De Finetti's representation of symmetric states in terms of i.i.d.\ 
states is, however, inherently limited to the case where $n \ll N$.
That is, given a large $N$-partite symmetric state, the i.i.d.\ 
property generally only holds approximatively for a small $n$-partite
subsystem~\cite{DiaFre80} (see Fig.~\ref{fig:deFin}), and the error in
the approximation is generally proportional to $\frac{n}{N}$.

To overcome this limitation, we propose a slightly relaxed variant of
the i.i.d.\ property where, roughly speaking, \emph{most}|but not
all|of the subsystems of a composite system are identical and
independent copies of each other. We then show the following
statement, extending de Finetti's representation theorem (see
Fig.~\ref{fig:deFin}): Given an $N$-partite quantum state $\rho^N$,
symmetry of $\rho^N$ implies that any $n$-partite part $\rho^n$ is
almost identical to a probabilistic mixture of states $\rho^n_\sigma$
that satisfy the relaxed i.i.d.\ property (with prototype $\sigma$),
as long as $n$ is slightly smaller than $N$ (e.g., $n \approx N -
\sqrt{N}$).

To make this more precise, consider a state $\rho^n$ on an $n$-partite
quantum system as well as a state $\sigma$ on a single subsystem. Then
$\rho^n$ is called \emph{$\binom{n}{m}$-i.i.d.\ (with prototype
  $\sigma$)} if it has the form $\sigma^{\otimes {m}} \otimes
\tilde{\rho}^{n-m}$, up to permutations of the subsystems, where
$\tilde{\rho}^{n-m}$ is an arbitrary state on $n-m$ subsystems. Note
that, for $m = n$, we retrieve the standard notion of i.i.d.\ states.
Our \emph{global\footnote{The term \emph{global} refers to the fact
    that the statement covers virtually the entire system (see
    Fig.~\ref{fig:deFin}).} representation theorem} can now be
formulated as follows (see Appendix~\ref{app:tech} for a more
technical statement and Appendix~\ref{app:proof} for a proof; see
also~\cite{Renner05} for a preliminary version as well
as~\cite{KoeMit07} for a nice generalisation of the result presented
here): Any $n$-partite part $\rho^n$ of an $N$-partite symmetric state
$\rho^N$ is approximated by a probabilistic mixture of states
$\rho^n_\sigma$ parameterised by $\sigma$, where each $\rho^n_\sigma$
is contained in the space spanned by $\binom{n}{n-r}$-i.i.d.\ states
with prototype $\sigma$, for $r \ll n$.  The error of the
approximation\footnote{The error is quantified in terms of the
  $L_1$-distance between operators. This distance measure, sometimes
  called \emph{trace distance}, is motivated by the fact that it
  corresponds to the probability of successfully distinguishing two
  quantum states.} is upper bounded by $\eps = 3 e^{- r \frac{N-n}{N}
  + d \ln (N-n) }$, where $d$ is the dimension of the subsystems,
i.e., the decrease is exponential in $r$.\footnote{If the subsystems
  are infinite-dimensional, $d$ can usually, for realistic systems, be
  substituted by some bound on the system's maximum energy.}  A
typical choice for the above parameters is $n := N - N^{\alpha}$ and
$r := N^\alpha$, where $\frac{1}{2} < \alpha < 1$.  Roughly speaking,
the global representation theorem then says that a symmetric state
$\rho^N$ can be seen as a mixture of i.i.d.\ states, as long as we
ignore $N^{\alpha}$ subsystems and, additionally, tolerate deviations
in at most $N^{\alpha}$ of the subsystems. (Note that $N^{\alpha}$ is
only sublinear in $N$ and the error $\eps$ decreases exponentially
fast in $N$.)

\section{Examples}

To get a feel for the above result, we have a look at some examples of
symmetric $N$-partite quantum states. For this, we assume that each
subsystem contains a set of $d$ mutually orthogonal (i.e., perfectly
distinguishable) states $\{\ket{0}, \ldots, \ket{d-1}\}$ (where $d=2$
in most examples).

\begin{enumerate}
\item \label{ex:mixt} Let $\rho^N$ be the uniform mixture of the two
  $N$-partite i.i.d.\ states $\ket{0}^{\otimes N}$ and
  $\ket{1}^{\otimes N}$. Obviously, any $n$-partite part $\rho^n$ of
  $\rho^N$, for $n \leq N$, still has the same structure, i.e., it is
  a convex combination of $\ket{0}^{\otimes n}$ and $\ket{1}^{\otimes
    n}$.  For this state, the representation theorem thus holds in a
  perfect sense (rather than only approximatively).  The example
  illustrates, however, that symmetric states (or parts of them) can
  generally not be approximated by \emph{one single} i.i.d.\ state
  $\sigma^{\otimes n}$, but only by mixtures of such states.
\item \label{ex:loc} Let $\rho^N$ be the uniform mixture of all states
  ${\ket{b_1} \otimes \cdots \otimes \ket{b_N}}$ where $(b_1, \cdots
  b_N) \in \{0,1\}^N$ are $N$-tuples of binary values with an even
  number of $1$s.  Any $n$-partite part $\rho^n$, for $n < N$, is
  equal to the i.i.d.\ state $\sigma^{\otimes n}$, where $\sigma$ is
  the uniform mixture of $\ket{0}$ and $\ket{1}$.  Note, however, that
  $\rho^N$ is not an i.i.d.\ state. This proves that the i.i.d.\ 
  property for an $N$-partite system cannot be verified by any
  experiment involving less than $N$ subsystems.
\item \label{ex:sup} Let $\rho^N$ be defined by the superposition
  (with equal amplitudes) of all $N$-partite states ${\ket{b_1}
    \otimes \cdots \otimes \ket{b_N}}$ with an even number of $1$s
  (note the difference to Example~\ref{ex:loc} where the state is
  defined by a mixture rather than a superposition of such states).
  While $\rho^N$ cannot be written as a mixture of i.i.d.\ states, it
  is easy to verify that any $n$-partite part $\rho^n$, for $n < N$,
  equals the uniform mixture of the two pure i.i.d.\ states
  $\ket{\bar{0}}^{\otimes n}$ and $\ket{\bar{1}}^{\otimes n}$, where
  $\ket{\bar{0}}:= \frac{1}{\sqrt{2}} (\ket{0} + \ket{1}$ and
  $\ket{\bar{1}} := \frac{1}{\sqrt{2}} (\ket{0} - \ket{1})$.
\item \label{ex:singlet} Let $N=2$ and let $\rho^2$ be the bipartite
  \emph{singlet} state defined by the antisymmetric vector
  $\frac{1}{\sqrt{2}} (\ket{0} \otimes \ket{1} - \ket{1} \otimes
  \ket{0})$. It is easy to verify that no (mixture of) i.i.d.\ states
  $\sigma^{\otimes 2}$ can have an overlap of more than $\frac{1}{4}$
  with $\rho^2$.  Because $\rho^2$ is symmetric\footnote{Note that,
    although the vector $\ket{\Psi^-} := {\frac{1}{\sqrt{2}} (\ket{0}
      \otimes \ket{1} - \ket{1} \otimes \ket{0})}$ defining the
    singlet is antisymmetric, i.e., $\pi \ket{\Psi^-} =
    -\ket{\Psi^-}$, the corresponding physical state $\rho^2:=
    \proj{\Psi^-}$ (represented as a density operator) is symmetric,
    i.e., $\pi \rho^2 \pi^\dagger = \rho^2$.} the example proves that
  symmetry is generally weaker than the i.i.d.\ 
  property. 
  In fact, our
  representation theorem does not yield any approximation in terms of
  i.i.d.\ states because the number of subsystems is small ($N=2$).
\item Let $\rho^N$ be the $N$-partite so-called \emph{cat state}
  defined by ${\frac{1}{\sqrt{2}} (\ket{0}^{\otimes N} +
    \ket{1}^{\otimes N})}$.  As in the above example, the overlap of
  any i.i.d.\ state $\sigma^{\otimes N}$ with $\rho^N$ is upper
  bounded by $\frac{1}{2}$, i.e., $\rho^N$ cannot be approximated by
  mixtures of i.i.d.\ states.  However, any $n$-partite part of
  $\rho^N$, for $n < N$, is exactly of the form of
  Example~\ref{ex:mixt}, i.e., a mixture of i.i.d.\ states.
\item \label{ex:antisym} Let $\rho^N$ be defined by the completely
  antisymmetric vector $\frac{1}{\sqrt{N!}}  \sum_{\pi} \sign(\pi)
  \cdot \pi(\ket{0} \otimes \ket{1} \otimes \cdots \otimes \ket{N-1})$
  with subsystems of dimension $d=N$. The state $\rho^N$ can be seen
  as a generalisation of the singlet state of Example~\ref{ex:singlet}
  (where $N=2$).  Although $\rho^N$ is symmetric, any bipartite part
  $\rho^2$ is a mixture of singlet states, and hence cannot be
  approximated by a mixture of i.i.d. states~\cite{CKMR06}.  The
  example thus illustrates that symmetry can only imply independence
  if the number $N$ of subsystems is sufficiently large compared to
  the dimension $d$ of the subsystems.
\item Let $\rho^N$ be the uniform mixture of all permutations of the
  $N$-partite state $\ket{0}^{\otimes N-1} \otimes \ket{1}$.
  Obviously, $\rho^N$ is an $\binom{N}{N-1}$-i.i.d.\ state with
  prototype $\ket{0}$.  However, the distance to any mixture of
  perfect i.i.d.\ states is at least $\frac{1}{2}$.  This implies that
  the $\binom{N}{N-r}$-i.i.d. property, for $r > 0$, is strictly
  weaker than the perfect i.i.d.\ property.
\end{enumerate}

\section{Applications}


Most physical measures that are used for the characterisation of large
composite systems (e.g., the energy or the temperature) are
\emph{robust} under disturbances of a small number of subsystems. In
particular, their values evaluated for a $\binom{N}{N-r}$-i.i.d.\ 
state $\rho^N$ with prototype $\sigma$ are approximated by their
values on the corresponding perfect i.i.d.\ state $\sigma^{\otimes
  N}$, as long as $r \ll N$. They are thus fully determined by the
prototype state $\sigma$ (which is the state of a single subsystem).
For example, if the measure $E$ is extensive (such as the energy or
the entropy) we have $E(\rho^N) \approx E(\sigma^{\otimes N}) = N
E(\sigma)$.  Furthermore, the prototype state $\sigma$ can be
determined by measurements applied to a limited number of subsystems.
Hence, under the assumption that the system's state $\rho^N$ is
$\binom{N}{N-r}$-i.i.d.\ for $r \ll N$, tomography is sufficient to
determine the value of any robust physical quantity.  The
representation theorem outlined in the previous section now implies
that the same is still true approximately under the sole assumption
that the system's state is symmetric.

A similar reasoning applies to problems in information theory and, in
particular, cryptography. A main challenge in these disciplines is to
characterise the resources (such as entanglement) which are needed to
perform certain tasks (e.g., teleportation). For this, it is often
convenient (and very common) to consider resources which consist of
many identical and independent parts or, more precisely, to assume
that the states describing the resources satisfy the i.i.d.\ property.
It is an immediate consequence of our representation theorem that this
assumption can be relaxed to a symmetry assumption.  This relaxation
is crucial because, in many information-theoretic scenarios, it
suffices to consider symmetric states in order to cover the most
general case. In fact, symmetry of the states can often be enforced by
applying randomly chosen permutations, as illustrated by the following
example (see also the Appendix~\ref{app:ex} for an additional
example).

\section{Example application: security of quantum key distribution}

\begin{figure} 
  \centering
  \includegraphics[scale=\picscale]{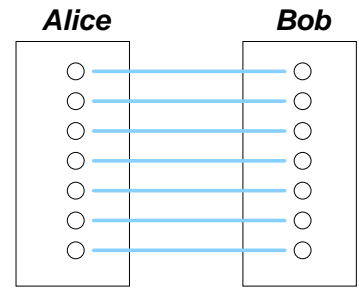}
\caption{In the first phase of a QKD scheme, called \emph{distribution
    phase}, Alice and Bob attempt to distribute a large number of
  entangled particle pairs, as depicted. In a subsequent
  \emph{distillation phase}, these are measured locally, resulting in
  a pair of raw keys held by Alice and Bob, respectively. The raw keys
  are then processed classically in order to produce a final secret
  key.
  \label{fig:qkd}}
\end{figure}

As indicated above, the global representation theorem has various
applications. As an example, we derive a generic result in quantum
cryptography~\cite{BenBra84,Ekert91}. The result implies security of a
large class of quantum key distribution (QKD) schemes against any
attack allowed by the laws of quantum physics. Generally speaking, QKD
is the art of distributing a (random) secret key to two distant
parties, using only communication over an insecure quantum channel as
well as an authentic\footnote{A communication channel is said to be
  \emph{authentic} if no adversary can alter the transmitted messages
  without being detected. Using a short initial key, an authentic
  channel can be simulated even if only a completely insecure channel
  is available~\cite{Stinso05}.} (but public) classical channel.
Typical QKD schemes consist of two subsequent
phases~\cite{Ekert91,BeBrMe92}: In a \emph{distribution phase}, one of
the parties, traditionally called \emph{Alice}, prepares $N$ entangled
particle pairs and sends one half of each pair over the quantum
channel to the other party, \emph{Bob} (cf.\ Fig.~\ref{fig:qkd}).
Then, in a \emph{distillation phase}, Alice and Bob apply local
measurements to their particles, resulting in a pair of correlated
classical strings, called \emph{raw keys}; finally, depending on an
estimate of the strength of correlation between their respective raw
keys, Alice and Bob employ some purely classical procedures to
transform them into identical secret keys.\footnote{The length of the
  generated keys depends on the correlation between the raw keys and
  might be zero if this correlation is too weak.}

As an adversary might tamper with the particles sent over the
(insecure) quantum channel, the joint state $\rho^{N}$ of the $N$
particle pairs held by Alice and Bob after the distribution phase is
generally (almost) arbitrary.  Hence, to prove security of the scheme
against general attacks, one has to show that the distillation phase
works correctly whatever the state $\rho^{N}$ is.  Because the space
of possible states $\rho^{N}$ is exponentially large in $N$, this
analysis is non-trivial and has only been possible for QKD protocols
which satisfy certain specific
requirements~\cite{Mayers96,LoCha99,ShoPre00}.\footnote{A typical
  requirement is that the protocol can be translated into a certain
  entanglement purification scheme~\cite{BBPSSW96}.} In fact, standard
information-theoretic arguments are usually restricted to situations
where the state $\rho^{N}$ is i.i.d., i.e., $\rho^{N} =
\sigma^{\otimes N}$. This, however, is only guaranteed for so-called
\emph{collective attacks}, where the adversary is bound to apply the
same operation separately to each of the particles sent over the
channel~\cite{BihMor97a,BBBGM02,DevWin05}.

Using the global representation theorem for symmetric states presented
in this article, it can be shown that security of a QKD scheme against
collective attacks implies security against arbitrary attacks (where
no restriction is imposed on the adversary).  The argument is based on
two observations: (i)~The security of the distillation phase only
depends on \emph{robust} properties of the state $\rho^{N}$ of the $N$
particle pairs held by Alice and Bob after the distribution phase,
i.e., security is not affected by alterations of a small number of
subsystems~\cite{Renner05}.  (ii)~If Alice and Bob both reorder their
particles according to a common randomly chosen permutation then the
resulting state of the particle pairs (averaged over all possible
permutations) is \emph{symmetric}.\footnote{Note that this holds even
  if the permutation is known to the adversary.} Now, given a QKD
scheme which is provably secure against collective attacks,
observation~(i) implies that the same scheme is secure whenever the
state $\rho^{N}$ has some $n$-partite part which is
$\binom{n}{n-r}$-i.i.d., where $N-n \ll N$ and $r \ll n$.  Hence, by
our representation theorem, it suffices to verify that $\rho^{N}$ is
symmetric, which is the case because of observation~(ii).  We thus get
the following result: If a QKD scheme is secure against collective
attacks then the same scheme, equipped with an additional randomised
permutation step inserted after the distribution
phase,\footnote{Inserting such a symmetrisation step is, however, only
  necessary if the scheme is not symmetric.  In fact, many schemes are
  already symmetric by construction (see, e.g., \cite{DEJMPS96}).}  is
secure against any attack allowed by the laws of quantum
physics.\footnote{This solves an open question originally raised by
  Biham and Mor~\cite{BihMor97a}.}

\section{Conclusions}

We have presented a de Finetti style representation theorem which
connects two properties that the physical state of a multi-partite
system can have: (i) \emph{symmetry:} swappings of the subsystems
leave the state unchanged; (ii) \emph{i.i.d.:} the individual parts of
the state are identical and mutually independent. The theorem states
that symmetry of a large system implies that the i.i.d.\ property
approximately holds on almost the entire system.

The i.i.d.\ property is often employed for the study of large systems,
but cannot usually be verified directly. In contrast, the symmetry
property is, for example, implied by the indistinguishability of the
subsystems or can be enforced by a random permutation. As the
representation theorem connects these two properties, it has
implications within various areas of physics (as does de Finetti's
original theorem, which is used, e.g., in mathematical physics and
statistical mechanics~\cite{FaSpVe80,FaLeVe88,RagWer89}). Furthermore,
the theorem has consequences for quantum information theory. For
instance, as demonstrated above, it implies that the security of a QKD
scheme against general attacks follows directly from its security
against collective attacks (which can be proved using standard
information-theoretic arguments).

The connection between symmetric and i.i.d.\ states is of particular
interest for foundational issues~\cite{Hudson81,CaFuSh02}. As
discussed above, it implies that symmetry suffices to predict physical
properties of a large quantum system given only data obtained from the
observation of a limited number of subsystems.  Since the ability to
make predictions is crucial in physics, one might go one step further
and postulate that a similar statement should be true within any
reasonable physical theory (other than quantum mechanics).  Such a
postulate would indeed restrict the space of possible theories.  For
example, within a theory where physical states are represented as
vectors in a real Hilbert space, even de Finetti's original
representation theorem cannot hold~\cite{CaFuSh02}.

\acknowledgments

I would like to thank Charles Bennett, Matthias Christandl, Artur
Ekert, Robert K\"onig, Ueli Maurer, and Graeme Mitchison for their
valuable and very helpful comments on earlier versions of this work.
This research is supported by HP Labs Bristol as well as by the
European Union through the Integrated Projects QAP (IST-3-015848),
SCALA (CT-015714), SECOQC and the QIP IRC (GR/S821176/01).

\appendix

\section{Technical statement of the representation theorem} \label{app:tech}

Let $\cH$ be a $d$-dimensional Hilbert space. The \emph{symmetric
  subspace of $\cH^{\otimes n}$}, denoted $\Sym^n(\cH)$, is the space
spanned by all vectors which are invariant under permutations of the
$n$~subsystems. Formally, let $S_n$ be the set of permutations on
$\{1, \ldots, n\}$. For any $\pi \in S_n$, we also write $\pi$ to
denote the unitary on $\cH^{\otimes n}$ which maps any product vector
$\phi_1 \otimes \cdots \otimes \phi_n$ to $\phi_{\pi^{-1}(1)} \otimes
\cdots \otimes \phi_{\pi^{-1}(n)}$.  Then $\Sym^n(\cH) := \{ \Psi \in
\cH^{\otimes n}: \, \pi \Psi = \Psi, \forall \pi \in S_n\}$.

Let $\nu$ be a rank-one projector on $\cH$.  A vector $\Psi \in
\cH^{\otimes n}$ is called \emph{$\binom{n}{m}$-i.i.d.}\ in $\nu$ if
there exists a permutation $\pi \in S_n$ such that $(\nu^{\otimes m}
\otimes \id^{\otimes n-m}) \pi \Psi = \pi \Psi$. Intuitively, this
means that the state defined by the vector $\Psi$ is of the form $\nu$
on (at least) $m$ subsystems.

Our main result establishes a connection between the symmetry and the
i.i.d.\ property as defined above.

\begin{theorem} \label{thm:main}
  Let $n, k, r \in \bbN$ and let $\cH$ be a $d$-dimensional Hilbert
  space. For any density operator $\rho^{n+k}$ on $\Sym^{n+k}(\cH)$
  there exists a measure $d \nu$ on the set $\cV$ of one-dimensional
  projectors on $\cH$ and a family of density operators $\rho^n_\nu$
  on $\Sym^n(\cH)$ such that, for any $\nu \in \cV$, $\rho^{n}_\nu$
  has support on the space spanned by all $\binom{n}{n-r}$-i.i.d.\ 
  vectors in $\nu$ and
  \[
    \bigl\| \tr_k(\rho^{n+k}) - \int \rho^n_\nu d \nu \bigr\|_1 
  \leq 
    3 e^{-\frac{k (r+1)}{n+k} + d \ln k} \ .
  \]
  Furthermore, if $\rho^{n+k}$ has rank one then the same is true for
  the operators $\rho^n_\nu$.
\end{theorem}

For any $N \in \bbN$, let $\rho^N$ be a density operator on
$\Sym^{N}(\cH)$ and let $\eps > 0$ be fixed. If we apply
Theorem~\ref{thm:main} with $k := \lceil\eps N\rceil $, $r :=
\lceil\eps N\rceil$, and $n:=N-k$, then the error in the approximation
provided by Theorem~\ref{thm:main} decreases exponentially fast in
$N$. Hence, very roughly speaking, the state $\rho^N$ is exponentially
(in $N$) close to a mixture of states which are i.i.d.\ except on an
arbitrarily small fraction (namely $r+k = 2 \eps N$) of the $N$
subsystems.

Note that a density operator $\rho^N$ on $\cH^{\otimes N}$ which is
symmetric under permutations, i.e., $\pi \rho^N \pi^{\dagger} =
\rho^N$ for any $\pi \in S_N$, cannot necessarily be seen as an
operator on $\Sym^N(\cH)$.\footnote{A simple example illustrating this
  fact is the operator $\rho^N = \frac{1}{d^N} \id_{\cH^{\otimes
      N}}$.}  Hence, in order to apply Theorem~\ref{thm:main} to
general symmetric quantum states, we need an additional lemma.  It
says that any permutation-invariant operator has a purification on a
symmetric subspace~\cite{Renner05,CKMR06}.


\begin{lemma} \label{lem:sympur}
  Let $\rho^N$ be a nonnegative operator on $\cH^{\otimes N}$ such
  that $\pi \rho^N \pi^{\dagger} = \rho^N$, for any $\pi \in S_N$.
  Then there exists a rank-one operator $\rhob^N$ on $\Sym^N(\cH
  \otimes \cK) \subseteq (\cH \otimes \cK)^{\otimes N}$, where $\cK
  \cong \cH$, such that $\rho^N = \tr_{\cK^{\otimes N}}(\rhob^N)$.
\end{lemma}

\section{Proof of the representation theorem} \label{app:proof}

The proof of Theorem~\ref{thm:main} is based on three technical lemmas
(Lemma~\ref{lem:gentle}--\ref{lem:symbasis}). The first can be seen as
a variant of Winter's gentle measurement lemma~\cite{Winter99} and is
implicitly used in related work~\cite{CKMR06}.

\begin{lemma} \label{lem:gentle}
  Let $\{\rho_\tau\}_{\tau \in \cT}$ be a family of nonnegative
  operators on a Hilbert space $\cH$ and let $\{P_\tau\}_{\tau \in
    \cT}$ be a family of projectors on $\cH$.  Then, for any measure
  $d \tau$ on $\cT$,
  \[
    \bigl\| \int (\rho_\tau 
    - P_\tau \rho_\tau P_\tau) d\tau \bigr\|_1 
  \leq 
    3 \bigl\| \int (\id-P_\tau) \rho_\tau d \tau \bigr\|_1 \ .
  \]
\end{lemma}

\begin{proof}
  Using the identity
  \begin{multline*}
    \rho_\tau - P_\tau \rho_\tau P_\tau \\
  =
    (\id-P_\tau) \rho_\tau + \rho_\tau (\id-P_\tau) - (\id-P_\tau) \rho_\tau (\id-P_\tau) \ ,
  \end{multline*}
  the triangle inequality, and the fact that $\| A \|_1 = \| A^\dagger
  \|_1$ holds for any operator $A$, we find
  \begin{equation} \label{eq:triabound}
    \bigl\| \int (\rho_\tau - P_\tau \rho_\tau P_\tau) d \tau \bigr\|_1
  \leq
    2 \alpha + \beta
  \end{equation}
  where 
  \begin{align*}
    \alpha 
  & := 
    \bigl\| \int (\id-P_\tau) \rho_\tau d\tau \bigr\|_1
\\
    \beta 
  & := 
    \bigl\| \int (\id-P_\tau) \rho_\tau (\id-P_\tau) d\tau \bigr\|_1 \ .
  \end{align*}
  Because, for any $\tau \in \cT$, the operator $(\id-P_\tau)
  \rho_\tau (\id-P_\tau)$ is nonnegative, the norm in the definition
  of $\beta$ can be replaced by a trace, that is,
  \begin{align*}
    \beta 
  & = 
    \tr\bigl( \int (\id-P_\tau) \rho_\tau (\id-P_\tau) d\tau \bigr) \\
  & =
    \tr\bigl( \int (\id-P_\tau) \rho_\tau d \tau \bigr) \ ,
  \end{align*}
  where the second equality follows from the cyclicity of the trace
  and the fact that $P_\tau P_\tau = P_\tau$.  Because $\tr(A) \leq \|
  A \|_1$ holds for any operator $A$, we conclude that $\beta \leq
  \alpha$. The statement then follows from~\eqref{eq:triabound}.
\end{proof}  

The next lemma is derived using basic arguments from representation
theory.

\begin{lemma} \label{lem:Schur}
  Let $A$ be an operator on $\cH^{\otimes n}$ and define
  \[
    \Gamma :=  \frac{\dim(\Sym^n(\cH))}{\tr(P_{\Sym^n(\cH)} A)} \int U^{\otimes n} A (U^\dagger)^{\otimes n} d U
  \]
  where $d U$ is the normalised Haar measure on the set of unitaries
  $\cU(\cH)$.  Then
  \[
    \Gamma P_{\Sym^n(\cH)}
  =
    P_{\Sym^n(\cH)} \ ,
  \]
  where $P_{\Sym^n(\cH)}$ is the projector onto the symmetric subspace
  of $\cH^{\otimes n}$.
\end{lemma}

\newcommand*{\rep}{\tau}

\begin{proof}
  The space $\cH^{\otimes n}$ can be decomposed into subspaces
  $\cH_\lambda$ labelled by Young diagrams $\lambda$ with $n$ boxes
  and at most $d := \dim(\cH)$ rows, i.e., $\cH^{\otimes n} \cong
  \bigoplus_\lambda \cH_\lambda^{\oplus m_\lambda}$, for some
  $m_\lambda \in \bbN$, such that the following holds. Let $\rep$ be
  the mapping from $\cU(\cH)$ to $\cH^{\otimes n}$ defined by $V
  \mapsto V^{\otimes n}$ and let $P_{\lambda}$ be the projector onto
  any of the subspaces $\cH_\lambda$ with Young diagram $\lambda$.
  Then $P_{\lambda}$ commutes with $\rep(V)$, for any $V \in
  \cU(\cH)$, and $V \mapsto P_{\lambda} \rep(V)$ is an irreducible
  representation of $\cU(\cH)$.  Furthermore, two such representations
  are equivalent if and only if their Young diagrams $\lambda$ are
  identical.
  
  Because $d U$ is the Haar measure, the operator $\Gamma$ commutes
  with $\rep(V)$, i.e., $\rep(V) \Gamma = \Gamma \rep(V)$, for any $V
  \in \cU(\cH)$. Let $P_\lambda$ and $P'_{\lambda'}$ be two projectors
  onto any of the subspaces $\cH_\lambda$ and $\cH_{\lambda'}$,
  respectively, as defined by the above decomposition.  Since these
  projectors commute with $\rep(V)$, we have
  \[
    (P_\lambda \rep(V)) (P_\lambda \Gamma P'_{\lambda'}) 
  =
    (P_ \lambda \Gamma P'_{\lambda'}) (P'_{\lambda'} \rep(V)) \ ,
  \]
  for all $V \in \cU(\cH)$. Consequently, by Schur's lemma, the
  operator $P_{\lambda} \Gamma P'_{\lambda'}$ acts like a scalar on
  $\cH_\lambda$ if $\lambda = \lambda'$ and equals zero otherwise. In
  particular, because for the Young diagram $\lambda = (n)$ with $n$
  boxes and one row $m_{\lambda} = 1$ and $P_{\lambda} =
  P_{\Sym^n(\cH)}$ holds, we find
  \begin{align*}
    \Gamma P_{\Sym^n(\cH)} 
  & =
    P_{\Sym^n(\cH)}  \Gamma P_{\Sym^n(\cH)} 
    + P_{\Sym^n(\cH)}^{\perp} \Gamma P_{\Sym^n(\cH)} \\
  & =
   \gamma \cdot P_{\Sym^n(\cH)} \ ,
  \end{align*}
  for some $\gamma \in \bbR$. Taking the trace on both sides of the
  equality gives $\gamma = 1$.
\end{proof}

\newcommand*{\projz}{P_{\ket{0}}}

Finally, we need an explicit basis of the symmetric subspace
$\Sym^n(\cH)$.

\begin{lemma} \label{lem:symbasis}
  Let $\cH$ be a $d$-dimensional Hilbert space with orthonormal basis
  $\{\omega_{b}\}_{b \in [d]}$, where $[d]:=\{1, \ldots, d\}$, and let
  $n \in \bbN$. Let $\Lambda^{n}_{d}$ be the set of $d$-tuples
  $\lambda = (\lambda_1, \ldots, \lambda_d) \in \bbN^d$ such that
  $\sum_{b=1}^d \lambda_b = n$, and, for any $\lambda \in
  \Lambda^n_d$, let
  $\cB_\lambda$ be the set of $n$-tuples $(b_1, \ldots, b_n) \in
  [d]^n$ such that $|\{i: b_i = b\}| = \lambda_b $ for $b \in [d]$.
  Then the family $\{\Phi_\lambda\}_{\lambda \in \Lambda^n_{d}}$ of
  vectors $\Phi_\lambda \in \cH^{\otimes n}$ defined by
  \[
    \Phi_\lambda
:=
  \sqrt{\frac{1}{|\cB_{\lambda}|}}
  \sum_{b \in \cB_\lambda}
     \omega_{b_1} \otimes \cdots \otimes \omega_{b_{n}}
  \]
  is an orthonormal basis of $\Sym^n(\cH)$. In particular,
  $\dim(\Sym^n(\cH)) = \| \Lambda^{n}_{d} \| = \binom{n+d-1}{n}$.
\end{lemma}

\begin{proof}
  See the standard literature on representation
  theory~\cite{FulHar91}.
\end{proof}

\begin{proof}[Proof of Theorem~\ref{thm:main}]
  Let $\nu_0$ be a fixed one-dimensional projector in $\cH$.  For $n,
  r \in \bbN$ and any unitary $U \in \cU(\cH)$, let $P^{n,r}_U$ be the
  projector onto the subspace of $\cH^{\otimes n}$ spanned by all
  $\binom{n}{n-r}$-i.i.d.\ vectors in $\nu := U \nu_0 U^\dagger$.  In
  particular,
  \begin{equation} \label{eq:Punitary}
    P^{n,r}_U = U^{\otimes n} P^{n,r}_\id (U^\dagger)^{\otimes n} \ .
  \end{equation}
  Define
\begin{align*}
  \rho^n_U 
& :=
  \dim(\Sym^k(\cH)) \cdot \tr_k\bigl(\id^{\otimes n} \otimes P^{k,0}_{U} 
  \cdot \rho^{n+k} \bigr)
 \\
  \rhob^n_U
& :=
  P^{n,r}_{U} \rho^n_U P^{n,r}_{U } \ ,
\end{align*}
where $\tr_k$ denotes the partial trace over the last $k$ subsystems,
and let $d U$ be the normalised Haar measure on $\cU(\cH)$. It is
straightforward to verify that $\rhob^n_U$ is a nonnegative operator.
Moreover, because $P^{k,0}_U$ and $\rho^{n+k}$ have rank one,
$\rhob^n_U$ has rank one as well. Since, by definition, $\rhob^n_U$
has support on the subspace containing all $\binom{n}{n-r}$-i.i.d.\ 
vectors in $\nu := U \nu_0 U^\dagger$, it suffices to show that
\begin{equation} \label{eq:deltabound}
  \delta
:=
  \bigl\| 
    \tr_k(\rho^{n+k}) - \int \rhob^n_U d U
  \bigr\|_1 
\leq
 3 e^{-\frac{k (r+1)}{n+k} + d \ln k} \ .
\end{equation}

Using~\eqref{eq:Punitary} together with the fact that $P^{k,0}_{\id}$
has support on the symmetric subspace $\Sym^k(\cH)$ and trace equal to
one, we can apply Lemma~\ref{lem:Schur} which gives
\[
   \dim(\Sym^k(\cH)) \int P^{k,0}_{U } d U \cdot  P_{\Sym^k(\cH)}
=
  P_{\Sym^k(\cH)} \ .
\]
Since, by assumption, $\rho^{n+k}$ has support on $\Sym^{n+k}(\cH)
\subseteq \cH^{\otimes n} \otimes \Sym^k(\cH)$, we conclude
\[
  \tr_k(\rho^{n+k}) 
=
   \int \rho^n_U d U
 \ .
\]
The distance $\delta$ can thus be rewritten as
\[
  \delta
=
   \bigl\| 
    \int (\rho^n_U - P^{n,r}_U \rho^n_U P^{n,r}_U) d U
  \bigr\|_1  \ .
\]
Let $(P^{n,r}_{U})^\perp := \id_{\cH^{\otimes n}} - P^{n,r}_U$ be the
projector orthogonal to $P^{n,r}_U$. By Lemma~\ref{lem:gentle}, we
have
\begin{align} 
  \delta
& \leq 
  3 \bigl\|
    \int (P^{n,r}_{U})^\perp \rho^n_U d U
  \bigr\|_1 \nonumber \\ \label{eq:deltaupbound}
& =
  3 \dim(\Sym^k(\cH)) \cdot  \bigl\| \tr_k ( \Gamma^{n+k} \rho^{n+k} ) \bigr\|_1
\end{align}
where
\[
  \Gamma^{n+k}
:= 
  \int (P^{n,r}_{U})^\perp \otimes P^{k,0}_{U} d U \ .
\]
Using again the fact that $\rho^{n+k}$ has support on
$\Sym^{n+k}(\cH)$ together with identity~\eqref{eq:Punitary} and
Lemma~\ref{lem:Schur}, the norm on the r.h.s.\ 
of~\eqref{eq:deltaupbound} can be rewritten as
\begin{multline*} 
  \bigl\| \tr_k ( \Gamma^{n+k} \rho^{n+k} ) \bigr\|_1
=
 \bigl\| \tr_k ( \Gamma^{n+k} P_{\Sym^{n+k}(\cH)} \rho^{n+k} ) \bigr\|_1  \\
=
 \gamma \cdot \tr(P_{\Sym^{n+k}(\cH)} \rho^{n+k})
= 
 \gamma \ ,
\end{multline*}
where 
\[
  \gamma
:=
  \frac{\tr\bigl(P_{\Sym^{n+k}(\cH)} (P^{n,r}_{\id})^\perp \otimes P^{k,0}_{\id} \bigr)}{\dim(\Sym^{n+k}(\cH))} \ .
\]
In order to show that~\eqref{eq:deltabound} holds, we insert this
into~\eqref{eq:deltaupbound}. Because $\dim(\Sym^k(\cH)) =
\binom{k+d-1}{k}$ (cf.\ Lemma~\ref{lem:symbasis}) and
$\binom{k+d-1}{k} \leq k^d$, for $k \geq 2$ (note that the statement
of the theorem is trivial for $k=1$), it remains to verify that
\begin{equation} \label{eq:gammaexpbound}
  \gamma \leq e^{-\frac{k (r+1)}{n+k}} \ .
\end{equation}

Let $\{\omega_b\}_{b \in [d]}$ be an orthonormal basis of $\cH$ such
that $\omega_b$, for $b=d$, is contained in the support of $\nu_0$.
Furthermore, let $\Lambda^{n+k}_d$ and $\{\Phi_{\lambda}\}_{\lambda
  \in \Lambda^{n+k}_d}$ be defined as in Lemma~\ref{lem:symbasis},
such that the latter is a basis of $\Sym^{n+k}(\cH)$. Then $\gamma$
can be rewritten as
\[
  \gamma
=
  \frac{1}{|\Lambda^{n+k}_d|}
    \sum_{\lambda \in \Lambda^{n+k}_d} \Phi_\lambda^\dagger 
    (P^{n,r}_{\id})^\perp \otimes P^{k,0}_{\id} \Phi_\lambda
  \ .
\]
A straightforward calculation shows that, for any $\lambda \in
\Lambda^{n+k}_d$ and $s := \sum_{b=1}^{d-1} \lambda_b$,
\[
  \Phi_\lambda^\dagger (P^{n,r}_{\id})^\perp \otimes P^{k,0}_{\id} \Phi_\lambda 
=
\begin{cases}
  0 & \text{if $s \leq r$} \\
  \frac{(n+k-s)! n!}{(n+k)! (n-s)!} & \text{otherwise.}
\end{cases}
\]
This immediately gives an upper bound on $\gamma$,
\[
  \gamma 
\leq
  \frac{(n+k-r-1)! n!}{(n+k)! (n-r-1)!} 
\leq
  \Bigl( \frac{n}{n+k} \Bigr)^{r+1} \ .
\]
Using the fact that, for any $\beta \in [0,1]$, $(1-\beta)^{1/\beta}
\leq e^{-1}$, we find, with $\beta := \frac{k}{n+k}$, 
\[
  \frac{n}{n+k}
=
  \bigl((1-\beta)^{1/\beta} \bigr)^\beta
\leq
  e^{-\frac{k}{n+k}} \ .
\]
This implies~\eqref{eq:gammaexpbound} and thus concludes the proof.
\end{proof}

\section{Evaluating extensive quantities on symmetric states} \label{app:ex}

In the following, we show that the global representation theorem
(Theorem~\ref{thm:main}) can be used to derive structural properties
of extensive quantities. In particular, we prove the following
proposition.

\begin{proposition}[Informal Proposition] \label{pr:add}
  Let $E$ be a concave extensive quantity which is continuous and
  robust (such that the variation of $E$ when altering $k$ subsystems
  is proportional to $k$).  Then, for any family of symmetric states
  $\rho^N$ on $\cH^{\otimes N}$ parameterised by $N \in \bbN$,
\begin{equation} 
    \lim_{N \to \infty} \frac{1}{N} E(\rho^N) \geq \min_{\sigma} E(\sigma)
 \ .
\end{equation}
\end{proposition}

Typical examples of quantities satisfying the assumptions of the
proposition are entropy measures, including the \emph{von Neumann
  entropy}, and entanglement measures (see also~\cite{Renner05}).

\begin{proof}[Proof sketch]
  Let $d := \dim(\cH)$, $k := N^{2/3}$, $r:=N^{2/3}$, and define
  $\rho^n := \tr_k(\rho^N)$, where $n = N-k$. According to
  Theorem~\ref{thm:main}, there exists a measure $d \sigma$ on the set
  of projectors on $\cH$ such that
  \[
    \bigl\| \rho^n - \int \rho^n_\sigma d \sigma \bigr\|_1
  \leq
    \delta := 3 e^{-N^{1/3} + d \ln(N)}
  \]
  where, for any $\sigma$, $\rho^n_\sigma$ has support on the space
  spanned by $\binom{n}{r}$-i.i.d.\ vectors in $\sigma$. Hence, using
  the continuity of $E$,
  \[
    E(\rho^n)
  \approx
    E\bigl(\int \rho^n_\sigma d \sigma \bigr) \ .
  \]
  Furthermore, by the concavity of $E$,
  \[
    E\bigl(\int \rho^n_\sigma d \sigma \bigr)
  \geq
    \int E(\rho^n_\sigma) d \sigma
  \geq
    \min_{\sigma} E(\rho^n_\sigma) \ .
  \]
  Combining this with the above and using the robustness of $E$, we
  find
  \[
    E(\rho^N)
  \approx
    E(\rho^n)
  \gtrapprox
    \min_{\sigma} E(\rho^n_\sigma) \ .
  \]
  Because $\rho^n_\sigma$ has support on the space of
  $\binom{n}{r}$-i.i.d.\ states in $\sigma$, robustness implies
  $E(\rho^n_\sigma) \approx E(\sigma^{\otimes n})$. The statement then
  follows from the fact that $E$ is extensive, i.e.,
  $E(\sigma^{\otimes n}) = n E(\sigma)$, and $\frac{n}{N} \approx 1$.
  \end{proof}
  
  Proposition~\ref{pr:add} provides some insights into a well-known
  problem of quantum information theory. Essentially, the problem is
  to prove the following conjecture, called \emph{additivity of the
    minimum output entropy of a quantum channel}~\cite{Shor04}.

  \begin{conjecture}
    For any trace-preserving completely positive map (CPM) $\cE$, the
    von Neumann entropy $S$ of the outcome of $\cE$, minimised over
    all possible inputs, is an extensive quantity.
  \end{conjecture}
  
  For a proof of this conjecture, it has to be shown that
\begin{equation} \label{eq:conj}
  \frac{1}{N} S(\cE^{\otimes N}(\rho^N)) 
\geq
  \min_{\sigma} S(\cE(\sigma)) 
\end{equation}
holds for any density operator $\rho^N$ on $\cH^{\otimes N}$. In the
special case where $\rho^N$ is symmetric, an asymptotic version
of~\eqref{eq:conj} follows from Proposition~\ref{pr:add}.  To see
this, it suffices to verify that the function $E$ defined by
$E(\rho^N) := S(\cE^{\otimes N}(\rho^N))$ satisfies the assumptions of
the proposition, which is straightforward.



\newcommand{\etalchar}[1]{$^{#1}$}

\end{document}